\documentclass[%
 reprint,
superscriptaddress,
bibnotes,
 amsmath,amssymb,
 aps,
]{revtex4-2}

\usepackage{graphicx}
\usepackage{dcolumn}
\usepackage{bm}
\usepackage{algorithm2e}

\usepackage{todonotes}

\usepackage{hyperref}
\usepackage{amsthm}
\usepackage{comment}
\usepackage{enumitem}
\usepackage{mathtools}
\usepackage{natbib}
\usepackage{amsfonts}
\usepackage{amssymb}
\usepackage{balance}

\newtheorem{prop}{Proposition}

\newtheorem*{rem}{Remark}

\newtheorem{result}{Result}

\newcommand{\tr}{\text{Tr}}
\newcommand{\mf}{\mathcal F}
\newcommand{\mg}{\mathcal G}
\newcommand{\me}{\mathcal E}
\newcommand{\mv}{\mathcal V}

\newcommand{\nei}{\partial}
\newcommand{\smin}{\setminus}

\usepackage {tikz}
\usetikzlibrary {positioning}

\tikzset{small dot/.style={fill=black,circle,scale=1.5}}
\tikzset{small dot 1/.style={fill=red!40,circle,scale=1.5}}
\tikzset{small dot 2/.style={fill=blue!40,circle,scale=1.5}}
\tikzset{small dot 3/.style={fill=green!40,circle,scale=1.5}}
\tikzset{small dot 4/.style={fill=red!70,circle,scale=1.5}}
\tikzset{small dot 5/.style={fill=blue!70,circle,scale=1.5}}
\tikzset{small dot 6/.style={fill=green!70,circle,scale=1.5}}
\tikzset{small dot 7/.style={fill=purple,circle,scale=1.5}}

\tikzset{small box/.style={fill=black,rectangle,scale=2}}
\tikzset{small box 1/.style={fill=red!40,rectangle,scale=2}}
\tikzset{small box 2/.style={fill=blue!40,rectangle,scale=2}}
\tikzset{small box 3/.style={fill=green!40,rectangle,scale=2}}
\tikzset{small box 4/.style={fill=red!70,rectangle,scale=2}}
\tikzset{small box 5/.style={fill=blue!70,rectangle,scale=2}}
\tikzset{small box 6/.style={fill=green!70,rectangle,scale=2}}

\tikzset{
    vertex/.style={circle, fill=gray!70, inner sep=2pt},
    edge/.style={thick, gray}
}

\usepackage{subcaption}
\usepackage{bbm}

\usepackage{lmodern}
\usepackage{physics}

\usepackage{pgfplots}
\usepackage{placeins}

\begin{document}

\title{Belief propagation for general graphical models with loops}

\author{Pedro Hack}
\email{Corresponding author. Email: pedro.hack@dlr.de}
\affiliation{Technical University of Munich, Germany}
\affiliation{German Aerospace Center, Germany}

\author{Jonas Hitter}
\email{jonas.hitter@tum.de}
\affiliation{German Aerospace Center, Germany}
\affiliation{Technical University of Munich, Germany}

\author{Christian B.~Mendl}
\email{christian.mendl@tum.de}
\affiliation{Technical University of Munich, Germany}

\author{Alexandru Paler}
\email{alexandru.paler@aalto.fi}
\affiliation{Aalto University, Finland}

\begin{abstract}
There is an increasing interest in scaling tensor network methods through belief propagation (BP), as well as increasing the accuracy of BP through tensor network methods. We develop a unification framework that takes an arbitrary graphical model with loops and provides message passing update rules and inference equations. We show that recent state-of-the-art methods regarding tensors and BP, like block belief propagation and tensor network message passing, are special instances of our framework. From a practical perspective, we discuss how our framework can be useful to understand the benefits of scheduling in BP, and show how it can be used for decoding purposes in quantum error correction. We simulate the computation of marginals, internal energy, Shannon entropy and the partition function on synthetic topologies (Kagome lattice and lattices resembling quantum error-correcting codes) and a real world topology of a power grid. The results show orders of magnitude accuracy increases for modest computational overheads. For the marginals, for example, we show that our framework can achieve an accuracy improvement of more than six orders of magnitude over tensor network BP.
\end{abstract}

\maketitle

\section{Introduction}

The development of generalized belief propagation (BP) protocols has received a great deal of attention recently. The interest started with the introduction of methods in the context of networks \cite{cantwell2019message,kirkley2021belief}, and continued with the incorporation of approximate tensor network contraction methods in order to propagate information \cite{wang2024tensor,piveteau2023tensor}. As a major result of this, there has been some advancement regarding the simulation of quantum systems and, hence, the situation in which one may achieve quantum supremacy \cite{beguvsic2024fast,tindall2024efficient}. The purpose of our work is to extend and unify the methods in these works. 

All the applications of interest described above can be formulated as inference tasks in the context of graphical models, which are tools typically used in the context of classical and quantum error correction ~\cite{mezard2009information,poulin2008iterative,bravyi2014efficient,iyer2015hardness,cao2022quantum,iolius2023decoding}. Hence, they can be treated as instances of the \textbf{decoding problem}~\cite{mezard2009information,iolius2023decoding}, where all the available information is implicitly contained in the graphical model and one ought to bring it forward via some computation in order to correct errors.
It is in this computation where the complexity issue known as the decoding problem lies. 

The most extended approach to the decoding problem is belief propagation~\cite{mezard2009information}, an algorithm introduced in cognitive sciences~\cite{pearl1988probabilistic} that allows to compute quantities that are relevant to decoding, like the partition function, internal energy, Shannon entropy or marginal over a few variables, in a computationally non-expensive way. 

Provided the graphical model under consideration is a tree, BP achieves exact results~\cite{mezard2009information,pearl1988probabilistic}. Despite the fact that BP has been empirically shown to work somewhat well for graphical models that locally resemble trees~\cite{frey1997revolution}, the issue remains that the exactness deteriorates as the graphical models becomes more loopy.

To obtain a method that improves on BP whenever it is not accurate, a natural approach is to try to better account correlations by computing them exactly, that is, to use more computational power in order to obtain accuracy. The first such attempt was named generalized BP~\cite{yedidia2000generalized}, an approach that was celebrated when it was introduced~\cite{yedidia2001bethe,yedidia2001characterization,yedidia2003understanding}, and that is still pursued, even within the quantum error correction community~\cite{old2023generalized}.

The main problem with generalized BP is that it does not provide a specific recipe, but rather a general approach that ought to be tuned to the concrete case in consideration. This was recently tackled~\cite{cantwell2019message,kirkley2021belief}, leading to the development of a method that, although following an idea similar to generalized BP, provides an explicit construction. Despite its success, however, the method has the drawback that it only applies to networks.

\subsection{Motivation}

Recently, and given the growing interest in both the use of message-passing schemes as general inference tools for tensor networks~\cite{alkabetz2021tensor,tindall2023gauging,guo2023block,pancotti2023one,beguvsic2024fast,tindall2024efficient} and the application of tensor networks to quantum error correction in general~\cite{ferris2014tensor,farrelly2021tensor,chubb2021statistical,cao2022quantum} and in \textbf{degenerate quantum maximum likelihood decoding}~\cite{bravyi2014efficient,iolius2023decoding}  and the \textbf{tensor network decoder}~\cite{piveteau2023tensor,gray2021hyper,beguvsic2024fast} in particular, approaches similar to the method of~\cite{kirkley2021belief,cantwell2019message} have been introduced in the context of tensor networks. An important recent contribution along these lines is \textbf{tensor network message passing}~\cite{wang2024tensor}, which was introduced in order to profit from BP when dealing with long loops and from tensor network contraction when dealing with short loops. These approaches follow the basic idea already in generalized BP in two different ways: 1) either they construct a single tensor network by grouping tensors together, like BlockBP does~\cite{guo2023block,kaufmann2024blockbp}, or, 2) they construct several directed graphs by grouping tensors together and combine the results given by them when doing inference. Here, we relate these two ways and extend them to arbitrary graphical models.

\begin{figure*}[!t]
    \includegraphics[width=\linewidth]{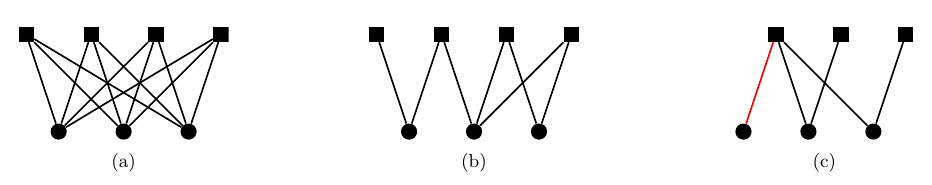}
    \caption{Graphical models: (a) general graphical model -- the degree of the nodes is arbitrary; (b) Network -- the degree of the factor nodes is always one or two; (c) Tensor network -- the degree of the variable nodes is always one or two. The squares are factors and the circles are variables. An edge is connecting a factor-variable pair whenever the factor depends on the variable. In the case of tensor networks, dangling edges (red) are associated with variable nodes of degree one.}
    \label{fig: different factor graphs}
\end{figure*}

\subsection{Graphical models}

Take a bipartite graph $\mathcal G = (\mathcal F \cup \mathcal V, \mathcal E)$ whose edges $\me$ join elements in $\mf$ with elements in $\mv$. A graphical model on $\mg$ is a function
    \begin{equation*}
        P(x_1,\dots,x_n) = \frac{1}{Z} \prod_{a \in \mathcal F} f_a(x_{\partial a}),
    \end{equation*}
    that is defined via a set of functions  
    $\{f_a\}_{a \in \mf}$, $f_a: \{x_i\}_{i \in \nei a} \to \mathbb C$ (also known as \textbf{factors}, potentials or \textbf{checks}), a set of variables $\{x_i\}_{i \in \mv}$ and a normalization constant or partition function
    \begin{equation*}
        Z = \sum_{x_1,\dots,x_n} \prod_{a \in \mf} f_a(x_{\nei a}).
    \end{equation*}
In particular, for all $a \in \mf$, the factor $f_a$ depends on $x_i$ provided there exists some $e \in \mathcal E$ joining them. We use the notation $\partial a$ for the set of variables that $f_a$ depends on. The notation $x_{\nei a}$ refers to a specific configuration of the variables.

In fact, we will indistinctly refer to $f_a$ as $a$ for all $a \in \mf$. Similarly, when dealing with a variable $x_i$, we will indistinctly refer to it by its associated node $i \in V$, and use the notation $\partial i$ for the set of functions that depend on $x_i$.

\begin{figure*}
    \includegraphics[width=\linewidth]{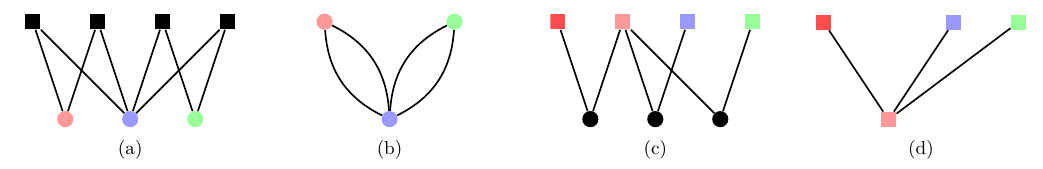}
    \caption{Factor graphs: (a) the factor graph of a network; (b) simplified graph of the network; (c) factor graph of a tensor network; (d) Simplified graph of the tensor network. Note that we have not merged double edges in the simplified network graphs. We do so since merging them does not affect the definition of the neighborhoods, and hence it does not modify the message passing method in any meaningful way.}
    \label{fig: simple fg}
\end{figure*}

We will use the following definitions and conventions:
\begin{enumerate}
    \item Given a graphical model $P$, we call its associated graph $\mg$ the \textbf{factor graph} or the \textbf{Tanner graph}~\cite{mezard2009information};
    \item all the considered graphical models are connected, such that the associated factor graph is connected; otherwise, we simply apply our analysis to each of its connected components;
    \item we assume all graphs to be undirected, unless explicitly stated otherwise;
    \item each variable $x_i$ takes values from the same finite set $X$ fulfilling $|X| \geq 2$ for simplicity;
    \item in the context of quantum error correction (QEC), we study the restricted case where $f_a: \{x_i\}_{i \in \nei a} \to \mathbb R_{\geq 0}$ for all $a \in \mf$, the so-called \textbf{probabilistic graphical models}~\cite{mezard2009information,koller2009probabilistic,pearl1988probabilistic};
    \item regarding the \textbf{visual} representation of graphical models \cite{mezard2009information}, we will use one \textbf{square} for each factor and one \textbf{circle} for each variable, connecting them by an edge whenever a factor depends on a variable (see Figure~\ref{fig: different factor graphs}). Provided a factor (variable) is connected to exactly two variables (factors), for simplicity, we may get rid of the square (circle) in the visual representation and directly join its nearest variables (factors) by an edge (see Figures~\ref{fig: simple fg}b and~\ref{fig: simple fg}d).
    \item given a graphical model the basic \textbf{quantities of interest} are: (variable and tensor) marginals, the internal energy, Shannon entropy and the partition function.
\end{enumerate}

As we illustrate in Figure~\ref{fig: different factor graphs}, we consider different classes of graphical models, and we show how to translate between them in SM Mapping graphical models to (tensor) networks:

\textbf{Networks}~\cite{kirkley2021belief} (also known as graphical models with pairwise potentials~\cite{yedidia2000generalized}) are graphical models whose functions $\{f_a\}_a$ depend at most on two variables, $|\partial a| \leq 2$ for all $a \in \mf$.

Since the functions that compose a network $P$ only depend on at most two variables, $P$ admits a simple representation via a graph $\mg = (\mv,\me)$, where we associate one node in $\mv$ to each variable in $P$ and one edge in $\me$ to each function in $P$. Moreover, a node is an endpoint of an edge whenever the function associated to the edge depends on the variable associated to the node. 

We call $\mg$ the \textbf{simplified graph associated to network $P$} (Figure~\ref{fig: simple fg}b). It will become clear later that, although one can avoid \textbf{double} edges, i.e. edges with the same  endpoints $e=(v_1,v_2)$ and $e'=(v_1,v_2)$ with $v_1,v_2 \in \mv$, by merging them together, this is not necessary and does not modify the algorithm in any meaningful way. 

In the following, we assume we are given a network whose functions depend exactly on two variables. Functions depending on a single variable can be naturally incorporated into the scheme~\cite{kirkley2021belief}.

\textbf{Tensor networks}~\cite{orus2014practical,bridgeman2017hand} are graphical models such that, for each variable $x_i$, there are at most two functions that depend on $x_i$, $|\partial i| \leq 2$ for all $i \in \mv$.

Since the variables that compose a tensor network $T= (T_i)_i$ are only shared by at most two functions, $T$ admits a simple representation via a graph $\mg = (\mv,\me)$, where we associate one node in $\mv$ to each function in $T$ and one edge in $\me$ to each variable in $T$. Moreover, a node is an endpoint of an edge whenever the function associated to the node depends on the variable associated to the edge. We call $\mg$ the \textbf{simplified graph associated to tensor network $T$} (Figure~\ref{fig: simple fg}d)

In the following, we assume we are given a tensor network whose variables are shared exactly by two tensors. If there were some variables such that a single tensor depends on them, the so-called \textbf{dangling edges} (Figure~\ref{fig: different factor graphs}c in red), these can be naturally incorporated into the scheme.

\begin{rem}
   While we focus here in non-negative graphical models, we consider a specific case of interest~\cite{alkabetz2021tensor,sahu2022efficient,tindall2023gauging,beguvsic2024fast} where $f_a: \{x_i\}_{i \in \nei a} \to \mathbb C$ in SM T-MITE for a double layer complex-valued tensor network. 
\end{rem}

\subsection{Neighborhoods and BP for networks with loops}

\begin{figure}[t!]
    \centering
    \includegraphics{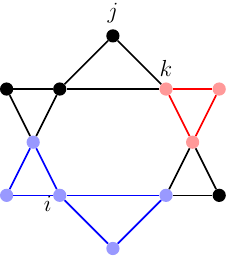}
    \caption{Neighborhoods definition. The \textbf{neighborhood around variable $i$} $N_i^{(\ell_0)}$ ($N_i^{(1)}$ in blue) includes node $i$ plus the edges that have it as endpoint the nearest neighbors of $i$ in $\mv$, together with the edges and nodes that belong to a path of length $\ell_0$ or less connecting two nearest neighbors of $i$. The \textbf{neighborhood difference from variable $i$ to $j$}, $N_{i \smin j}^{(\ell_0)}$ ($N_{k \smin j}^{(3)}$ in red), consists of $i$ together with all the edges that belong to $N_i^{\ell_0}$ and are not included in $N_j^{\ell_0}$ plus the nodes at their ends.}
    \label{fig: graph example}
\end{figure}

In order to introduce their BP equations on a network, the authors of~\cite{kirkley2021belief} define \textbf{neighborhoods}, that are constructed given some integer parameter $\ell_0 \geq 0$ and some variables $i,j \in \mv$. Figure~\ref{fig: graph example} describes the types of neighborhoods, $N_i^{\ell_0}$ and $N_{i \setminus j}^{\ell_0}$, which effectively are subgraphs of the simplified graph $\mg$.

We call $\ell_0$ the \textbf{loop-parameter}. Once we choose the value of this parameter, we use it to construct all the neighborhoods. In what follows, for simplicity, we may drop the parameter $\ell_0$ and simply refer to the neighborhoods as $N_i$ or $N_{i \setminus j}$. Using these neighborhoods, Kirkley et al.~\cite{kirkley2021belief} proposed a variation of BP

This method uses a family of messages defined on the finite set $X$ which are uniformly initialized:
\begin{equation*}
    \begin{split}
    &\{m_{i \to j}^{(t)} \}_{i \in V, j \in N_i, t \geq 0}, \text{ where } m_{i \to j}^{(t)}: X \to \mathbb R_{\geq 0},\\
    &m_{i \to j}^{(0)} (x_i) \equiv 1/|X| \text{ for all } x_i \in X,
    \end{split}
\end{equation*}

The messages are updated, for $t \geq 0$ and for all $x_i \in X$, according to the following equation:
\begin{equation}
    \label{eq: update network}
    m_{i \to j}^{(t+1)} (x_i) \propto
    \tr_{\smin \{ x_i \}} \left( \prod_{a \in N_{i \smin j}} f_a \prod_{k \in N_{i \smin j}} m_{k \to i}^{(t)} \right),
\end{equation}
where we use the compact tensor notation $\tr_{\setminus A}( \cdot )$~\cite{orus2014practical} for the \textbf{trace} over the variables outside of a set $A$ in order to avoid explicitly writing the variables in the sum. We also us $\propto$ in this and the following equations to indicate we omit a normalization constant that ensures $m_{i \to j}^{(t+1)}(\cdot)$ is a probability distribution. In general, the normalization of messages in BP schemes is not an actual theoretical necessity, but a practical addition to avoid numerical underflow~\cite{pearl1988probabilistic,weiss2000correctness}. Preventing this will justify the appearance of normalization constants in all the update equations that follow.

The updates are executed for a given number of iterations, or until the equations have converged. After that, Kirkley et al.~\cite{kirkley2021belief} provide inference equations to compute the quantities of interest. Importantly, whenever the \textbf{loop bound is fulfilled} or there are only \textbf{bounded loops}, that is, the length of the loops in $\mg$ is bounded by $\ell_0+2$, the inference equations are exact. This is the case since, whenever the bound holds, we can use the neighborhoods to construct a tree equivalent to the original network. If this does not hold, the update equations are still well defined, although they are no longer exact. Still, these equations have been reported to give good results in this context~\cite{kirkley2021belief} provided the graph is \textbf{locally dense and globally sparse}.

\subsection{Building a tree-equivalent graph}

Let us take an arbitrary $\mg$, which might contain loops. We can use the following algorithm to associate a tree-equivalent graph starting from a single \textbf{seed} node. The algorithm is using the neighborhoods as defined in Figure~\ref{fig: graph example}, and is parameterized by an integer $\ell_0 \geq 0$:
\begin{enumerate}[label=(P\roman*)]
    \item Pick randomly a seed $i_0 \in \mv$, and let $A=\{i_0\}$.
    \item While there exist vertices in $N_{i_0} \setminus A$ that are connected to vertices outside of $N_{i_0}$, pick randomly any such a vertex $i \in N_{i_0} \setminus A$ and add it to $A$.
    \item Take as $i$ the first element in $A$ such that there exist vertices in $N_{i \setminus \alpha(i)} \setminus A$ that are connected to vertices outside of $N_{i \setminus \alpha(i)}$, where $\alpha(i)$ is the \textbf{ancestor} of $i$ in $A$ (that is, the element through which $i$ was added to $A$), pick randomly any such that a vertex $i \in N_{i \smin \alpha(i)} \setminus A$ and add it to $A$.
    \item Repeat (Piii) until there is no $i \in A$ that can add nodes to $A$ via (Piii).
\end{enumerate}

The output of this algorithm is the \textbf{tree-equivalent graph} $\mg_{TE} = (\mv_{TE}, \me_{TE})$, where
\begin{equation}
    \label{eq: vertices eq tree}
    \mv_{TE} \equiv \{N_{i_0}\} \cup \{N_{i \setminus \alpha(i)}\}_{i \in A \setminus \{i_0\}}  
\end{equation}
and whose edges $\me_{TE}$ connect $N_{i_0}$ with $N_{i \setminus \alpha(i)}$ provided $\alpha(i)=i_0$ and $N_{i \setminus \alpha(i)}$ with $N_{k \setminus \alpha(k)}$ provided $\alpha(i)=k$. 

\begin{rem}
This algorithm is grouping nodes according to parameter $\ell_0$. In case $\ell_0$ is smaller than the largest loop in $\mg$, the output $\mg_{TE}$ will not be a tree. Otherwise, the output will be a tree. This observation is used in the Results.
\end{rem}

\begin{rem}
Given some $i \in A$, we denote by $D(i)$ the \textbf{descendants} of $i$, that is, the set of $j \in A$ such that $\alpha(j)=i$. The descendants will be used in the Results.
\end{rem}

\subsection{Contribution}

First, we introduce the tree-equivalent method (TE) and its inference equations, an approach for networks that follows the graph decomposition that is implicitly existing in the BP for loopy networks. TE allows us to generalize our approach and to apply it to any topology. 

Second, we extend both the BP on loopy networks and TE to arbitrary graphical models in the Methods. This allows us to improve on tensor network message passing by providing an algorithm that works for any tensor network and obtaining inference equations for global quantities. Our generalized tensor network message passing method (see Results) is useful for degenerate quantum maximum likelihood decoding~\cite{bravyi2014efficient,kaufmann2024blockbp} as well as non-degenerate decoding (see Applications).

Finally, we present numerical evidence for the performance of our method in the Results, and show that for particular quantities of interest, on realistic as well as synthetic benchmark models, our approach achieves up to seven orders of magnitude accuracy improvements compared to standard BP for a comparatively moderate computational cost.

\begin{table}[!t]
\centering
    \begin{tabular}{c c  c c c} 
        Method                                  & Graphical Model  \\
        \hline
        \textbf{Tree-equivalent TE}    & Arbitrary                   \\
        \textbf{Multiple intersecting TE (MITE)}        & Arbitrary                   \\ 
        BP for loopy networks (N-MITE)      & Network           \\
        TNMP (T-MITE)        & Tensor network        \\
        Block BP (TE)~     & Tensor network  \\
        \hline
    \end{tabular}
     \caption{Different approaches in this paper. Our contributions are bolded, and the other methods are instances of our MITE generalization.}
     \label{tbl:overview}
\end{table}

\section{Methods}

Herein we describe a generalization of the neighborhoods-method to any graphical model. Before we build our generalization, we introduce a simple version,
where one is message passing on a tree-equivalent graph obtained by the algorithm from the Introduction. This allows us to gain insights into how to unify the different message passing methods on arbitrary models (see Results). An overview of our general methods and discussions with other BP methods are listed in Table~\ref{tbl:overview}.

\subsection{A tree-equivalent approach to networks}

One can run on the tree-equivalent graph a message passing algorithm similar to the one from~\cite{kirkley2021belief,cantwell2019message}. We will call this the \textbf{tree-equivalent} (TE) method:
\begin{enumerate}
    \item We introduce variables for the neighborhoods of $\mg_{TE}$, effectively relabeling the vertices in Eq.~\eqref{eq: vertices eq tree}, 
        \begin{enumerate}[label=(\roman*)]
            \item denoting $N_{i_0}$ by $i_0$ 
            \item denoting $N_{i \setminus \alpha(i)}$ by $i$.
        \end{enumerate}

    \item We define the set of messages
        \begin{equation*}
            \begin{split}
                &\{m_{i \to \alpha(i)}^{(t)}, m_{\alpha(i) \to i}^{(t)}\}_{i \in A, t \geq 0}, \text{  where}\\
                &m_{i \to \alpha(i)}^{(t)}, m_{\alpha(i) \to i}^{(t)} : X \to \mathbb R_{\geq 0}.        
            \end{split}
        \end{equation*}

    \item We initialize the messages uniformly and we update them, for $t \geq 0$, according to Equations~\ref{no norm I}. The latter are defined for all $x_i \in X$, where $\{S_{i \smin \alpha(i)}\}_{i \in A}$ denotes the product of functions included in either $N_{i \smin \alpha(i)}$, provided $i \in A \smin \{ i_0 \}$, or $N_{i_0}$, if $i=i_0$, $\alpha^2(i) \equiv \alpha(\alpha(i))$ for all $i \in \mv$, and we have omitted a normalization constant in each equation.

    \item Once the messages have converged, we can use the inference equations in SM The tree-equivalent method.
\end{enumerate}

\begin{widetext}
\begin{equation}
\label{no norm I}
\begin{split}
   &m_{i \to \alpha(i)}^{(t+1)} (x_i) 
   \propto 
   \tr \left( S_{i \smin \alpha(i)} \prod_{k \in D(i)} m_{k \to i}^{(t)} \right),\\
   &m_{\alpha(i) \to i}^{(t+1)} (x_i) \propto
   \tr \left( S_{\alpha(i) \smin \alpha^2(i)} m_{\alpha^2(i) \to \alpha(i)}^{(t)} \prod_{k \in D(\alpha(i)) \smin \{i\}} m_{k \to \alpha(i)}^{(t)} \right).
   \end{split}
\end{equation}
\end{widetext}

The inference equations from SM The tree-equivalent method will be exact whenever $\mg_{TE}$ is a tree, and will not be exact whenever $\mg_{TE}$ is not a tree. In the  non-tree case are adapted versions of those used on trees: Instead of using $\{S_{i \setminus \alpha(i)}\}_{i \in A}$ in the update equations, we use $\{S_{i_m \setminus \cup_{n=1}^{m-1} i_n}\}_{m \geq 0}$, where $A=(i_m)_{m \geq 0}$ is a numbering of $A$ according to the order in which they are incorporated following the algorithm from the Introduction \footnote{In this instance, some messages will have dimension larger than $|X|$.}.
In the non-tree case, we expect the inference equations to be accurate for locally dense and globally sparse graphs~\cite{kirkley2021belief} (See Results).

\subsection{BP for loopy graphical models}

At this stage we can present our \textbf{multiple intersecting tree equivalent} (MITE) method, which is a generalization of message-passing on arbitrary graphical models (not only tree-equivalent graphs). A direct and practical result is that we can translate the method from the Introduction, which we call \textbf{network MITE} (N-MITE) in the following, to tensor networks and obtain \textbf{tensor MITE} (T-MITE). The latter will be benchmarked in the Results.

The intuition behind MITE is to treat variables and factors in the same way when defining neighborhoods, and to only distinguish them when introducing the messages. This means we will treat factors in the spirit of tensors from the tensor network approach (SM BP for loopy tensor networks), and variables in the spirit of variables from the network approach (Introduction). Our generalization allows us to show that, for example, BlockBP is an instance of TE, and TNMP is an instance of T-MITE.

We start by considering a graphical model $P$ and its associated factor graph $\mg = (\mf \cup \mv, \me)$. For each vertex $i \in \mf \cup \mv$, we create a neighborhood $S_i$ as before and define $S_{i \setminus j}$ and $S_{i \cap j \to i}$ in the same way. We treat the factor graph of a general graphical model as the simplified graph of a tensor network.

\subsubsection{Bounded loops}
\label{sec:bpgm_bound}

In case $\mg$ fulfills the loop bound, we:
\begin{enumerate}
    \item Define the set of messages
\begin{equation}
\label{eq: bounded gm def}
    \begin{split}
       &\{m_{i \to j}^{(t)} \}_{i \in \mf \cup \mv, j \in N_i, t \geq 0}, \text{ where}\\
        &m_{i \to j}^{(t)}: X^{|X_{i \setminus j}|} \to \mathbb R_{\geq 0} \text{ if } i \in \mf \text{ and}\\
       &m_{i \to j}^{(t)}: X \to \mathbb R_{\geq 0} \text{ if } i \in \mv,
    \end{split}
\end{equation}
with $X_{i \setminus j}$ denoting the set of variables that factor $i$ depends on and that do not belong to to $S_j$.
\item We initialize the messages uniformly,
and update them, for $t \geq 0$ according to Equation~\ref{no norm II}.
\item Once the messages have converged, we can use the equations in SM The inference equations for arbitrary graphical models with bounded loops to perform inference.
\end{enumerate}

\begin{equation}
\label{no norm II}
\begin{cases}
m_{i \to j}^{(t+1)} (x_i) \propto
\tr_{\setminus x_i} ( S_{i \setminus j} \prod_{k \in S_{i \setminus j} } m_{k \to i}^{(t)}) \\

\text{ for all } x_{i} \in X \text{ if } i \in \mv,\\
\\

m_{i \to j}^{(t+1)} (x_{i \setminus j}) \propto 
\tr_{\setminus x_{i \setminus j}} ( S_{i \setminus j} \prod_{k \in S_{i \setminus j} } m_{k \to i}^{(t)}) \\

\text{ for all } x_{i \setminus j} \in X_{i \setminus j} \text{ if } i \in \mf.
\end{cases}
\end{equation}

\subsubsection{Unbounded loops}
\label{sec:bpgm_unbound}

Whenever the loop length is not bounded by the loop-parameter, we take care of the missing legs issue (SM BP for loopy tensor networks and Figure S1) by adding \textbf{intersection messages}. In this case, we:
\begin{enumerate}
    \item Define the set of messages
        \begin{equation}
            \begin{split}
         &\{m_{i \to j}^{(t)},  m^{(t)}_{i \cap j \to i}\}_{i \in \mf \cup \mv, j \in N_i, t \geq 0}, \text{ where}\\
         &\{m_{i \to j}^{(t)}\}_{i \in \mf \cup \mv, j \in N_i, t \geq 0} \text{ is defined as in Eq.}~\eqref{eq: bounded gm def} \text{ and}\\
         &m_{i \cap j \to i}^{(t)}: X^{|X_{i \cap j \to i}|} \to \mathbb R_{\geq 0},        
            \end{split}
        \end{equation}
 with $X_{i \cap j \to i}$ 
 denoting the set of variables that $m_{i \to j}^{(t)}$ does not depend upon and that are connected to functions in $S_{i \cap j \to i}$.  In the following, in case $|X_{i \cap j \to i}| = 0$, we assume $m^{(t)}_{i \cap j \to i}$ to have dimension one and to be equal to one for all $t \geq 0$.
 
     \item We initialize the messages uniformly, and update them, for $t \geq 0$, according to Equation~\ref{eq: update gm}.
     
     \item Once the messages have converged, we can use the equations from the bounded case with modifications analogous to those in the tensor network case. In case the graphical model is a (tensor) network, the update equations in Eq.~\eqref{eq: update gm} reduce to Eq.~\eqref{eq: update network} (those in the SM).
\end{enumerate}

\begin{equation}
\label{eq: update gm}
\begin{cases}
    m_{i \to j}^{(t+1)} (x_i) \propto 
    \tr_{\setminus x_i} ( S_{i \setminus j} m_{i \cap j \to i}^{(t)} \prod_{k \in S_{i \setminus j} } m_{k \to i}^{(t)}) \\
    \text{ for all } x_{i} \in X \text{ if } i \in \mv,\\
    \\

    m_{i \to j}^{(t+1)} (x_{i \setminus j}) \propto 
    \tr_{\setminus x_{i \setminus j}} ( S_{i \setminus j} m_{i \cap j \to i}^{(t)} \prod_{k \in S_{i \setminus j} } m_{k \to i}^{(t)}) \\
    \text{ for all } x_{i \smin j} \in X_{i \smin j} \text{ if } i \in \mf, \\
    \\
    
    m_{i \cap j \to i}^{(t+1)} (x_{i \cap j \to i}) \propto
    \tr_{\setminus x_{i \cap j \to i}} ( S_{i \cap j \to i} m_{i \to j}^{(t)} 
    \prod_{k \in S_{i \cap j \to i} } m_{k \to i}^{(t)}) \\
    \text{ for all } x_{i \cap j \to i} \in X_{i \cap j \to i}.
\end{cases}
\end{equation}

\section{Results}

Our results are two-fold: a) we provide a theoretical unification to arbitrary graphical models among different existing loopy BP methods (BlockBP, TNMP, our TE and MITE methods); b) we demonstrate numerically our general message passing algorithm on graphical models with loop sizes greater than a chosen $\ell_0$.

\subsection{Unification of message passing algorithms}

The development of TE and MITE allows us to unify the state-of-the-art loopy BP methods. This includes:
\begin{enumerate}
    \item Gaining insight into N-MITE.
    \item Showing that tensor network message passing (TNMP)~\cite{wang2024tensor} is an instance of MITE and, hence, incorporating inference equations for global quantities into TNMP, and 
    making it available for quantum degenerate maximum likelihood decoding, for example.
    \item Showing that BlockBP~\cite{alkabetz2021tensor,guo2023block,kaufmann2024blockbp} is an instance of TE.
\end{enumerate}

\subsubsection{TE vs N-MITE}

TE can explain why N-MITE is exact only when the loop bound is fulfilled, and also why it is a good choice whenever the loop bound is not fulfilled.

We can think of the method of~\cite{kirkley2021belief} as creating a tree-equivalent directed graph: the edge between each pairs of nodes in that tree is directed from $N_{j \smin \alpha(j)}$ to $N_{\alpha(j) \smin \alpha^2(j)}$. The edge direction means that N-MITE sends only a single message along the edge. Once the messages have converged, N-MITE averages the inference results given by each of the trees to obtain the quantities of interest. If the loop bound is fulfilled,  then all trees are exact and averaging over them has no effect. However, if the loop bound is not fulfilled, then averaging over them is a sensible choice. In fact, while one may think that N-MITE allows to reduce the computational complexity while keeping the accuracy intact, this is not necessarily the case (SM The tree-equivalent method):
\begin{enumerate}
    \item when there are only \textbf{bounded loops}, one can find for most graphs $\mg$ a seed such that either the tree-equivalent method is less complex than N-MITE, or vice versa;
    \item when there are \textbf{unbounded loops} (i.e. the loop bound is not fulfilled), it may be difficult to predict which seed node to use in the tree-equivalent method in order to achieve the best trade-off between accuracy and complexity.
\end{enumerate}

In the Applications we discuss how the relation between TE and N-MITE can provide some justification to the improvement on BP given by the scheduling of messages.

\subsubsection{TNMP is N-MITE with different neighborhoods}

Although TNMP introduces neighborhood in the context of tensor networks, there is no difference if we do so in the context of networks, and then compare the neighborhoods directly to the ones from N-MITE. Crucially, the key difference between TNMP~\cite{wang2024tensor} and N-MITE~\cite{kirkley2021belief} is the definition of the neighborhoods.

The TNMP-method starts from a network which is turned into a tensor network, and then an algorithm called \textbf{tensor network message passing} is essentially passing messages between variables in a manner similar to N-MITE. The TNMP-method is tailored to tensor networks associated with statistical mechanics. This prevents TNMP to being applied to general tensor networks, in contrast with T-MITE. A second shortcoming in~\cite{wang2024tensor} is that no equation to infer a global quantity is provided~\footnote{Their local inference equations are in~\cite[Equations (S7) and (S8)]{wang2024tensor}.}

We will refer to the neighborhoods in N-MITE as N-MITE-neighborhood (and denote them by $N_{i}^{K,\ell_0}$) and to those in the TNMP-method as TNMP-neighborhoods (and denote them by $N_{i}^{W,\ell_0}$). We consider the  general relation between TNMP- and N-MITE in SM The relation between N-MITE and TNMP.

Given the simplified graph of a network $\mg$, we define TNMP-neighborhoods using a parameter that plays a role similar to that of the loop-parameter $\ell_0$. Given a subset of the network $N \subseteq \mg$, the \textbf{outer distance of $N$}, $\text{min}_{(a,b)} d_{a,b} (\partial N)$, is the length of the smallest path outside of $N$ that connects two variables in the boundary of $N$, $a,b \in \partial N$. The boundary of $N$ refers to variables that have a function that depends on them and is not contained in $N$.

Given some variable $i \in \mg$ and an integer parameter $\ell_0 \geq 0$, the TNMP-neighborhood $N_i^{W,\ell_0}$ is constructed by incorporating variables and functions into $N_i^{W,\ell_0}$ until $\text{min}_{(a,b)} d_{a,b} ( \partial N_i^{W,\ell_0}) \geq \ell_0$. In fact~\cite{wang2024tensor}, one can construct $N_i^{W,\ell_0}$ by recursively adding variables and edges as follows:
\begin{enumerate}[label=(R\roman*)]
    \item Add $i$ as well as all the edges connected to it and the variables they connect it to.
    \item For each $\ell \leq \ell_0$ starting from $\ell=1$, add all variables and edges along paths of length $\ell$ or less that connect a pair of variables in the boundary $N_i^{W,\ell_0}$. Repeat the procedure until neither edges nor variables can be added anymore. After that, start the step again, this time with $\ell+1$ (provided $\ell +1 \leq \ell_0$).
\end{enumerate}

We refer to the iteration at which each node $v$ (and analogously for each edge) is incorporated to $N_i^{W,\ell_0}$ as its \textbf{generation} and denote it by $g_i(v)$ or, whenever $i$ is clear, $g(v)$.

By comparing the neighborhoods in each method, we get the following:

\begin{result}
MITE can be applied to TNMP-neighborhoods. Moreover, we can use the MITE inference equations on TNMP-neighborhoods.
\end{result}

To see this, let us argue as in the context N-MITE-neighborhoods. We start by assuming that the neighborhoods we use $\{N_i^{W,\ell_0}\}_{i \in \mv}$ take into account all the correlations in the graph, 
$\text{min}_{(a,b)} d_{a,b} ( \partial N_i^{W,\ell_0}) = \infty$. If that is the case, then one can show (see the proof of (Wvii) in SM The relation between N-MITE and TNMP) that there exists some integer $t_{\ell_0} \geq 0$ such that 
\begin{equation*}
N_i^{W,\ell_0} = N_i^{K,t_{\ell_0}}    
\end{equation*}
for all $i \in \mv$.
Hence, one is under the assumptions where one can derive exact inference equations via N-MITE-neighborhoods and, since the neighborhoods are equal, the same equations also hold for TNMP-neighborhoods.
This allows us to go beyond~\cite{wang2024tensor} and provide inference equations for global quantities.

If the assumption that $\text{min}_{(a,b)} d_{a,b} (\partial N_i^{W,\ell_0}) = \infty$ fails, we can again use the equations derived assuming it holds and do approximate inference using them (see SM The relation between N-MITE and TNMP).

\subsubsection{BlockBP is the tree-equivalent approach}

In BlockBP, a PEPS tensor network~\cite{guo2023block,orus2014practical,bridgeman2017hand} is partitioned into blocks of square shape and messages are exchanged between blocks in the spirit of the tree-equivalent method. BlockBP's inference performance is better than simply doing tensor network BP, because the correlations within the blocks are computed with a large accuracy. The accuracy comes at the cost of increased computational complexity, and approximate contraction methods like bMPS have also been considered~\cite{guo2023block,kaufmann2024blockbp}.

\begin{result}
BlockBP, and the tree-equivalent method in general, is a limit case of T-MITE, in the sense that only variables belong to the intersection between different neighborhoods.
\end{result}

Although the inference equations from SM BP for loopy tensor networks may not be well-defined, they can be naturally extended to this limit case. For example, we can consider the inference equation for the partition function used in~\cite{kaufmann2024blockbp}, which we derive (see SM Derivation of the partition function for the tree-equivalent approach) along the lines of our derivation of the partition function in T-MITE (see SM The inference equations for tensor networks).

\begin{figure*}[!t]
    \includegraphics[width=\linewidth]{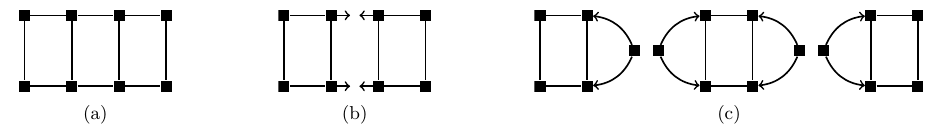}
    \caption{An improvement on the BlockBP: a) Original graph; b) an instance of BlockBP; c) an instance of an N-MITE-like algorithm applied on the original graph.}
    \label{fig: blockbp kirkley}
\end{figure*}

Although one can improve on BlockBP in some instances where the algorithm is exact (see Figure~\ref{fig: blockbp kirkley} for example), it is in general not possible to do so when the blocks form a two-dimensional tree. We discuss the relation between the two methods whenever BlockBP is not exact in the Discussion.

\subsection{Numerical experiments}
\label{sec:numerics}

Our theoretical, generalized approach to BP is also practically relevant. To illustrate this, we implemented T-MITE with Python 3.12 and Quimb 1.11.2 \cite{gray2018quimb}, and ran simulations on a standard i7-1370P CPU with 16GB RAM.

\begin{figure*}
    \includegraphics[width=\linewidth]{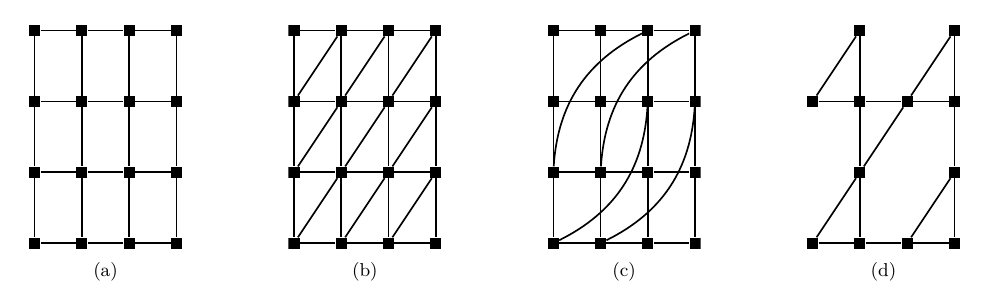}
    \caption{Synthetic tensor networks used to benchmark our generalization: (a) the square-$d_0$ lattice; (b) the square-$d_1$ lattice; (c) the square-$d_2$ lattice; (d) The Kagome lattice. We show the 4x4 instance for each of them.}
    \label{fig: benchmark graphs}
\end{figure*}

\begin{figure}[!tb]
\centering
    \includegraphics[width=\columnwidth]{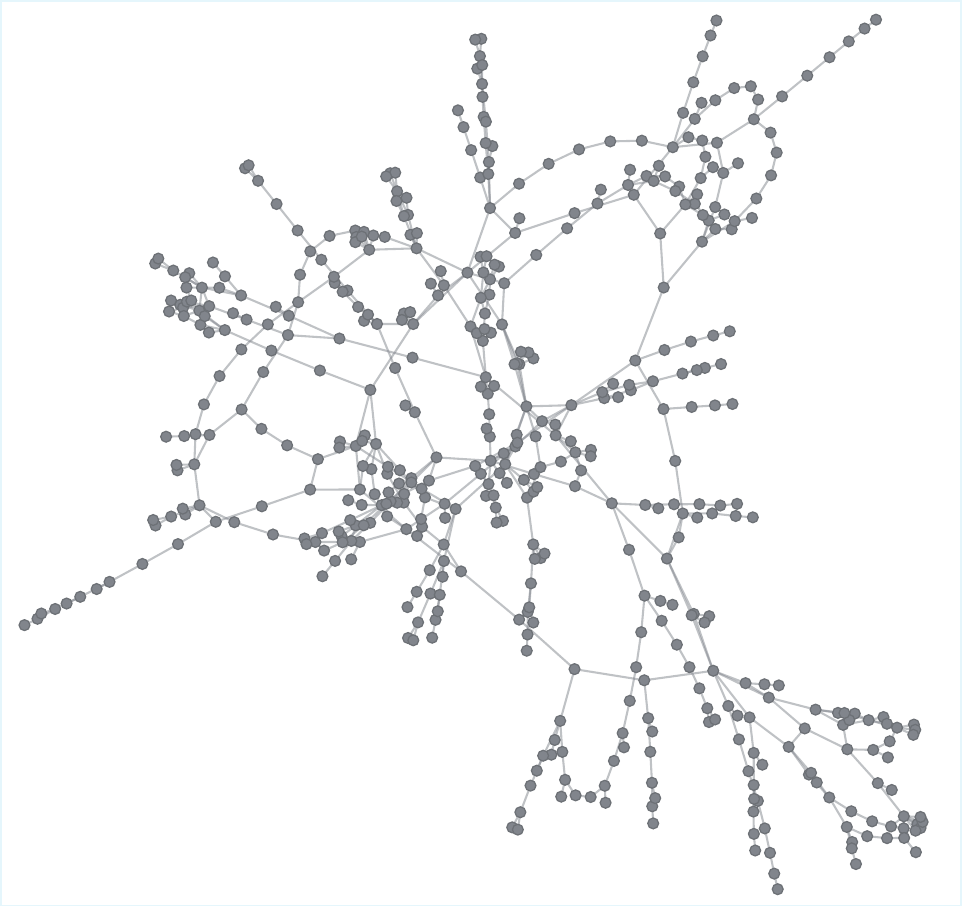}
    \caption{The power lattice represents a real power grid.}
    \label{fig: power toplogy}
\end{figure}

We benchmarked the method using five different tensor network topologies (four synthetic ones and a real one):
\begin{enumerate}
    \item The \textbf{square-$d_0$ lattice} is a square lattice (Figure~\ref{fig: benchmark graphs}a). This topology is closely related to the \textbf{surface quantum error-correcting code}~\cite{iolius2023decoding} (see Applications).
    
    \item The \textbf{square-$d_1$ lattice} is a square lattice with right diagonal connections of distance one (Figure~\ref{fig: benchmark graphs}b).
    
    \item The \textbf{square-$d_2$ lattice} is a square lattice with right diagonal connections of distance two (Figure~\ref{fig: benchmark graphs}c). This lattice is related to the \textbf{bivariate bicycle code}~\cite{bravyi2024high} (see Applications).
    
    \item The \textbf{Kagome lattice} (Figure~\ref{fig: benchmark graphs}d) has applications in material science and in the study of \textbf{frustrated spin systems} \cite{yin2022topological}.
    
    \item The \textbf{power lattice} represents the structure of an electrical power grid~\cite{davis2011university} (see Figure \ref{fig: power toplogy}). This lattice is important because it is not synthetic and it was used as a benchmark network for 
    N-MITE~\cite{kirkley2021belief}.
\end{enumerate}

For each of the above topologies, the values that the tensors $T_i^{\beta}$ in the tensor network $ T^{\beta} = (T_i^{\beta})_i$ take are given by the \textbf{Ising model}, that is, we use $X=\{0,1\}$ and
\begin{equation}
    \label{ising tensors}
    T_i^{\beta}(x_1,\dots,x_{m_i}) \equiv
    \begin{cases}
        e^{-\beta} \text{ if } x_1= \cdots = x_{m_i} = 1,\\
        1 \text{ otherwise,}
    \end{cases}
\end{equation}
where $\beta$ is the \textbf{inverse temperature}. We simulate $T^{\beta}$ for inverse temperature values $\beta = 4/k$ in the range $k=1,\dots,16$.

We are interested in observing how the value of $\ell_0$ is influencing the accuracy of the inferred quantities of interest. At the same time, we compare the runtime of our approach with the standard belief propagation approach. Effectively, our numeric simulations answer the following questions:
\begin{enumerate}
    \item How large are the accuracies achievable with our generalized message passing, as a function of $\ell_0$?
    \item How large is the computational overhead in order to achieve orders of magnitude accuracy increases?
\end{enumerate}

To answer the first question, for each topology and inverse temperature, we perform the simulations for different values of $\ell_0$ (e.g. $\ell_0 =0, 1, \ldots$). Each simulation is executed for a maximum number of 20 message passing iterations, or until the distance between the message at one iteration and the previous one is below $10^{-12}$.

\begin{figure}[h]
    \centering
    \includegraphics[width=0.8\columnwidth]{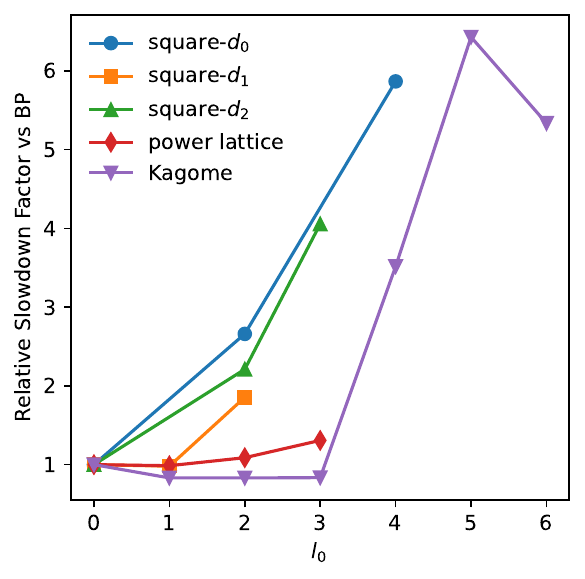}
    \caption{Runtime relative to tensor network belief propagation, i.e. T-MITE with $\ell_0=0$, for the different topologies and $\ell_0$ values we have considered. The reported runtimes are an average over the different values of the inverse temperature $\beta$ we have studied.}
    \label{fig: runtime}
\end{figure}

\begin{figure*}
    \centering
    \includegraphics[width=\textwidth]{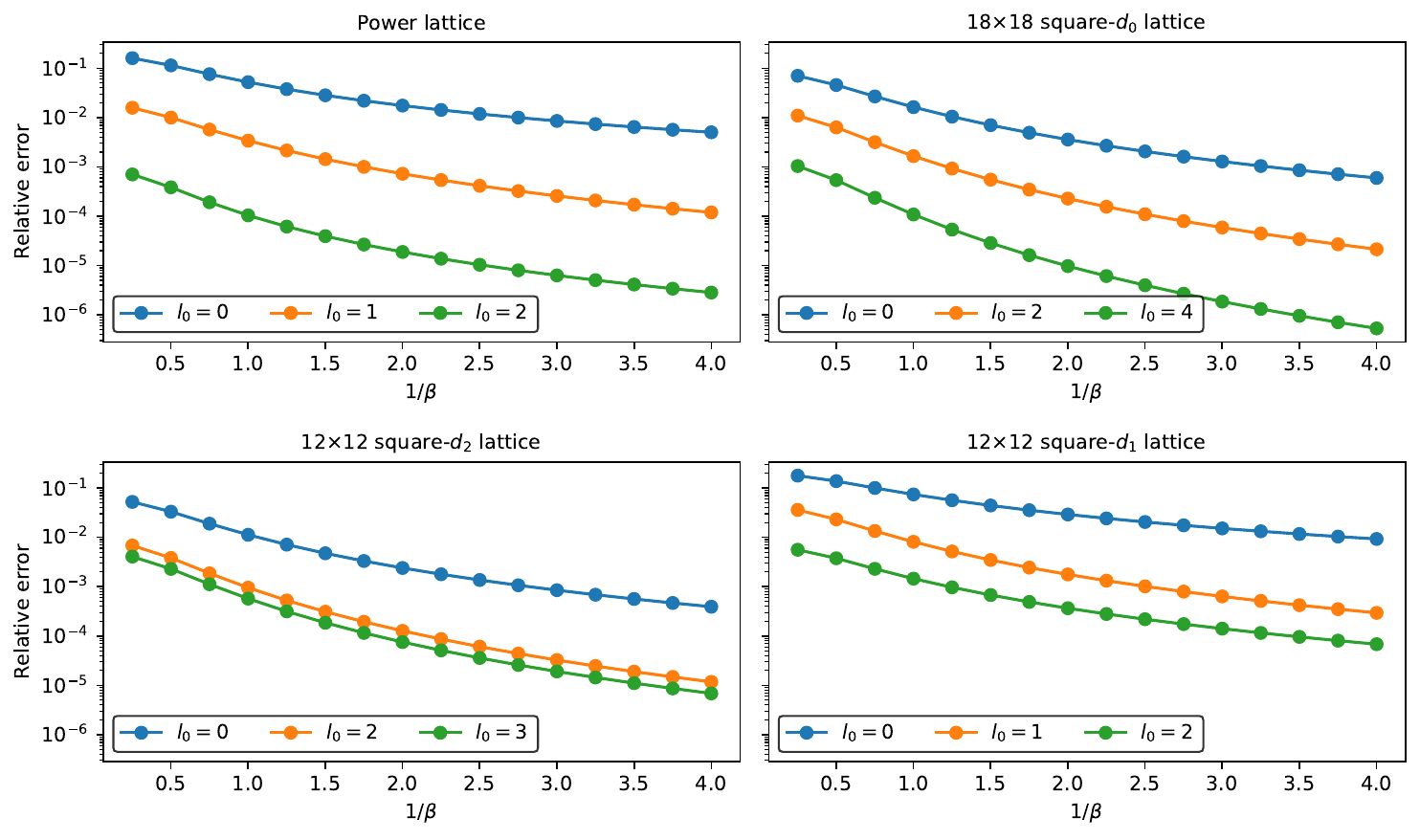}
    \caption{Mean percentage of error in the tensor marginals. The $\ell_0$ values which are not included, and which lie between the maximal and minimal ones included, coincide with an included value. For instance, in the square-$d_0$ lattice, the algorithm instance with $\ell_0=1$ coincides with that with $\ell_0=0$.}
    \label{fig: marginals results}
\end{figure*}

We can show that, for the investigated topologies and values of $\ell_0$, the computational overhead is very small. Figure~\ref{fig: runtime} shows that, for example, for the square-$d_0$ lattice there is at most a factor six slowdown for $\ell_0=4$, while the relative accuracy improvement (Figure~\ref{fig: marginals results}) ranges between three and five orders of magnitude.

We illustrate a single quantity of interest, the mean percentage of error in the tensor marginals. This relative error is the average over $i$ of the percentage of error in the marginal distribution $p_{T_i}$, for the square-$d_0$, square-$d_1$, square-$d_2$ and power lattices in Figure~\ref{fig: marginals results}. We show the marginal results for the Kagome Lattice, together with some numerical results for the partition function, entropy and internal energy in the SM Monotonicity in $\ell_0$.  

We delay some of the numerical results to SM Monotonicity in $\ell_0$ in order to discuss the \textbf{monotonicity} in $\ell_0$ of T-MITE, that is, whether the accuracy and runtime increase monotonically with the parameter $\ell_0$.

\begin{rem}[Monotonicity in $\ell_0$]
Increased values of $\ell_0$ do not necessarily imply improved accuracy (e.g. this is hinted by the relative accuracies for $\ell_0=2$ and $\ell_0=3$ being very close for the square-$d_2$ lattice in Figure~\ref{fig: marginals results}), or increased runtime (e.g. Kagome lattice for $\ell_0=5$ and $\ell_0=6$ in Figure~\ref{fig: marginals results}).
\end{rem}

We have emphasized during the text that our algorithms are tailored for locally dense and globally sparse topologies.  This is the context where all the benchmarks in~\cite{kirkley2021belief} take place, and this also explains why a lack of monotonicity in $\ell_0$ was not observed. Nevertheless, we have benchmarked T-MITE in general and application-oriented topologies.

We conjecture, that the monotonicity remark is enabled by our benchmark topologies having a \emph{hierarchy of loops}, with most loops being local and short and only a few being global and long.

\begin{rem}[Loop structure within lattices]
The square-$d_0$, -$d_1$ and -$d_2$ lattice do not present a hierarchy of loops. The power lattice (also used in~\cite{kirkley2021belief}) presents a two-level hierarchy since it is a real topology. The Kagome lattice interpolates between the two scenarios, since it has a very regular structure at the highest level, which is composed of hexagonal substructures that themselves can be decomposed into triangles.
\end{rem}

\section{Applications}

We discuss in this section two applications of our work: a) the improvement of scheduling on BP and b) the use of tensor networks and T-MITE for quantum error correction.

\subsection{Improving BP decoding through scheduling}

Despite it having been reported in some instances, the improvements provided by scheduling, at least in the context of QEC~\cite{du2023layered,chang2008lower, panteleev2021degenerate, kuo2020refined}, are not well understood. 

Scheduling is a variation of the original BP scheme where messages are not flooded, that is, they are not sent through all the edges of the QEC Tanner graph at every time step.  Instead, there is a schedule that determines which parts of the Tanner graph exchange information with others at every time step. 

\begin{result}
N-MITE is a theoretical construction that introduces message scheduling and improves the performance of BP.
\end{result}

\begin{figure}[!t]
    \includegraphics[width=\columnwidth]{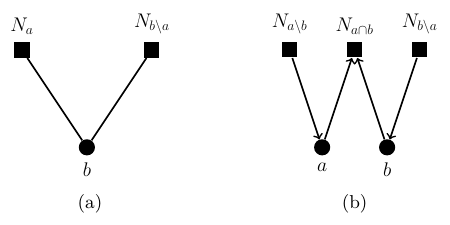}
    \caption{Tanner graphs: a) the tree-equivalent method (i.e. BP without scheduling) and b) N-MITE (i.e. BP with scheduling).}
    \label{fig:scheduling}
\end{figure}

For simplicity, we focus our discussion on networks. In order to understand how message scheduling improvement is related to N-MITE, let us consider a graph $\mg$ such that $\mg = N_a \cup N_b$ for $a,b \in \mv$ and some loop-parameter $\ell_0$. The Tanner graph for the tree-equivalent (flooded BP) and N-MITE (scheduled BP) are in Figure~\ref{fig:scheduling}. 

On the one hand, in the case of message flooding BP, taking $a$ as seed in the tree-equivalent method, then $m_{a \to b}$ communicates all the information in $N_a$ to $N_{b \smin a}$ and $m_{b \to a}$ communicates all the information in  $N_{b \smin a}$ to $N_a$.

On the other hand, N-MITE provides an instance where the improvement given by scheduling can be clearly explained. N-MITE profits from the fact that $\mg = N_{a \cap b} \cup N_{a \smin b} \cup N_{b \smin a}$ to get $m_{a \to b}$ to communicate all the information in $N_{a \smin b}$ to $N_{a \cap b}$ and $m_{b \to a}$ to communicate all the information in $N_{b \smin a}$ to $N_{a \cap b}$. This splits the information transmission between $a$ and $b$ in two stages, since it is only after the update process has finished, i.e. at the inference stage, that the information in $N_{b \smin a}$ reaches $N_a$ and the information in $N_{a \smin b}$ reaches $N_b$.

Consequently, we can think of N-MITE as the tree-equivalent method with scheduling, and we improve on the performance of BP.
In order to achieve the improvement, one needs to pick the seed for the tree-equivalent method following the proof of (Tii) in SM The tree-equivalent method.

\subsection{BP for loopy tensor networks in quantum error correction}

The most extended use of tensor networks for quantum error correction is in the context of \textbf{degenerate} quantum maximum likelihood decoding. In there, the tensor network decoder is used to compute the probabilities of the different error cosets, that is, tensor networks are built to compute the sum of the probabilities of all errors that are equivalent up to some stabilizer~\cite{bravyi2014efficient}. While the application of the tools developed here to this scenario is straightforward, we can also use them in the context of \textbf{non-degenerate} quantum maximum likelihood decoding (e.g. minimum weight perfect matching decoding~\cite{iolius2023decoding}).

\begin{figure}
    \includegraphics[width=\columnwidth]{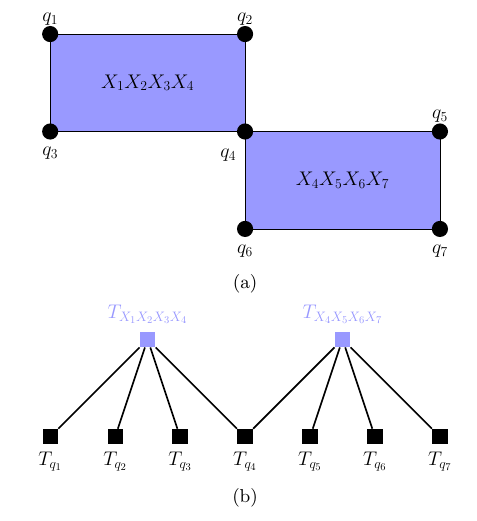}
    \caption{a) Excerpt of the surface code with one data qubit at each node and one syndrome qubit (associated to an $X$ stabilizer acting on the qubits at the nodes) in each plaquette; b) Extended tensor network associated to the excerpt,  after mapping the factor graph to a tensor network where the tensors in the upper part are indicator functions that are one whenever the combination of their variables yields the observed syndrome, and those in the lower part are diagonal tensors containing the noise model.}
    \label{fig: surface code tn}
\end{figure}

When performing \textbf{non-degenerate} quantum maximum likelihood decoding in this scenario, since the surface code is a CSS code, it is customary to run BP on two Tanner graphs, often called also \emph{un-correlated decoding}: one corresponding to $X$ stabilizers and another one to their $Z$ counterparts~\cite{iolius2023decoding}. As one can see in Figure~\ref{fig: surface code tn}, these Tanner graphs are tensor networks, whose constituents are indicator functions whose output is a one provided the combination of inputs matches the observed syndrome. However, they lack the noise model if they are directly used. Hence, one should use an \textbf{extended} tensor network, which includes, for each variable $i \in \mv$, a diagonal tensor $T_i$ whose entry $T_i(x,x)$ corresponds to the a priori probability of qubit $i$ undergoing Pauli error $x$. One can then use any of the BP tensor network methods developed here in order to infer the marginal distribution over each variable and decide what error correction operation to apply.

\section{Discussion}

We have developed a unified view regarding the algorithms that have been proposed in order to make generalized BP explicit.  We have considered the tree-equivalent and MITE methods in the context of networks, tensor networks and general graphical models. We derived BP schemes and provided inference equations. Concretely, we related BlockBP to the MITE by introducing the tree-equivalent approach, which is applicable to any graph, in contrast to how tailored BlockBP is to the lattice. 

Moreover, aside from showing how to extend the TNMP-method to arbitrary tensor networks, we have shown that the method essentially relies on the same approximation as N-MITE, allowing us to make it available for degenerate quantum maximum likelihood decoding. Lastly, our extension of N-MITE to arbitrary factor graphs makes it available for non-degenerate quantum maximum likelihood decoding in general.  

Our methods and results can be used as a principled approach for understanding the accuracy-complexity trade off at the intersection of message-passing and tensor networks methods. This is closely related to the decomposition of graphs into loop hierarchies and the monotonicity in $\ell_0$. Additionally, we could complement our work with heuristics that allow us, provided the loop bound is not fulfilled, to pick good candidates among all the tree-equivalent methods and to compare how such candidates perform compared to MITE.

Our key insight is that the construction of neighborhoods and the exchange of non-intersection messages between the neighborhoods play a role in the accuracy-complexity trade-off of QEC decoding. From a practical perspective, the goal would be to obtain codes which are locally dense and globally sparse, such that their decoding would be improved. To this end, it is worthwhile to construct QEC codes guided by MITE and the numerical evidence coming from the usual BP on loopy graphs.

Finally, future work will focus on practical improvements of BP decoders by exploiting the connection between message scheduling algorithms in BP and MITE. Moreover, it remains a challenge to prvide data regarding the decoding capabilities of MITE in the context of QEC.

\section*{Acknowledgements}

\subsection*{Funding}
P. Hack acknowledges funding by the Munich Quantum Valley. The research is part of the Munich Quantum Valley, which is supported by the Bavarian state government with funds from the Hightech Agenda Bayern Plus.

This research was developed in part with funding from the Defense Advanced Research Projects Agency [under the Quantum Benchmarking (QB) program under award no. HR00112230006 and HR001121S0026 contracts], and was supported by the QuantERA grant EQUIP through the Academy of Finland, decision number 352188. The views, opinions and/or findings expressed are those of the author(s) and should not be interpreted as representing the official views or policies of the Department of Defense or the U.S. Government.

\subsection*{Author contributions}

The conceptualization was carried out by P.H., C.B.M., and A.P. The investigation was performed by P.H., and A.P. Methodology was developed by P.H. The original draft was written by P.H., C.B.M., and A.P. P.H., C.B.M., and A.P. contributed to reviewing and editing the manuscript.  Validation involved P.H., J.H., and A.P. Formal analysis was conducted by P.H. Software was developed by P.H., and J.H. Visualization was prepared by P.H., and A.P. Resources were provided by P.H. Funding was acquired by P.H., C.B.M., and A.P. Supervision was provided by P.H., C.B.M., and A.P. Project administration was handled by P.H., and A.P.

\subsection*{Competing interests}

All authors declare that they have no competing interests.

\subsection*{Data and Materials Availability}

All data needed to evaluate and reproduce the results in the paper are present in the paper and/or the Supplementary Materials. The source code is open sourced at \url{https://github.com/alexandrupaler/bpgenloop}.

\balance

\bibliography{__main}

\clearpage

\setcounter{figure}{0}
\renewcommand{\figurename}{FIG.}
\renewcommand{\thefigure}{S\arabic{figure}}

\end{document}